\documentclass[journal]{IEEEtran}
\usepackage{amsmath,amssymb}
\usepackage{graphicx}
\usepackage{cite}
\usepackage{xcolor}
\usepackage{booktabs}
\usepackage[hidelinks]{hyperref}

\begin{document}

\title{Delayed Convergence of S-Parameter Sensitivities to Design Parameters in FDTD Simulations}

\author{Byungkwan~Kim,~\IEEEmembership{Member,~IEEE}%
\thanks{This work has been submitted to the IEEE for possible publication. Copyright may be transferred without notice, after which this version may no longer be accessible.}%
\thanks{B. Kim is with the Department of Radio and Information Communications Engineering, Chungnam National University, Daejeon,
Korea (e-mail: byungkwan.kim@cnu.ac.kr).}%
\thanks{This work was supported in part by the ICT R\&D program of
MSIT/IITP (RS-2023-00228599, Key Source Technology Development of
Intelligent Microwave Energy System for ISM Applications) and in part
by the National Research Foundation of Korea (NRF) grant funded by the
Korea government (MSIT) (No.~RS-2026-25519181).}}

\maketitle

\begin{abstract}
In the finite-difference time-domain (FDTD) method, the sensitivities of S-parameters to design parameters come from discrete Fourier transforms of
responses ending at a simulation time $T$. We show that, when a design parameter moves a resonance, the sensitivity converges
later than the S-parameter. For a single mode with field decay time $\tau$, the truncation error of the sensitivity is
$(1+T/\tau)e^{-T/\tau}$, against $e^{-T/\tau}$ for the resonant part of the S-parameter, so a 1\,\% sensitivity error needs $T\ge6.6\tau$
instead of $4.6\tau$. Automatic differentiation, the discrete adjoint and finite differences on the same run carry this error, so
their agreement cannot reveal it. In three solvers, with no fitted parameter, the estimate predicts the
simulated error within 4\,\% at cavity resonances with $Q$ of 1000--7000; at lower $Q$ the error is larger. On a cavity whose slowest
mode is weakly coupled, common end criteria leave $|S_{11}|$ within 0.5\,dB of a 1.3\,dB dip but its sensitivity 10--72\,\% in error. A
matrix-pencil extrapolation inside the differentiable computation recovers the sensitivity on a cavity with field probes from a run 66
times shorter. In a single-pole filter loop with a symmetric objective, truncation leaves the maximum of the objective in place, and the
optimizer reaches the resonance of a full run despite the wrong gradient. With two coupled poles, the truncated objective moves the
optimum: on runs of $1\tau$ it over-couples by $\Delta k\approx1.24/Q$, and an equal-ripple objective still shifts it by about
$0.2\bar\alpha$ at $3\tau$, but not at $4.6\tau$.

\end{abstract}


\section{Introduction}\label{sec:intro}
\IEEEPARstart{G}{radient-based} design of microwave components with time-domain solvers needs the sensitivities of the
S-parameters to the design parameters. Adjoint variable methods obtain them from one extra simulation~\cite{nikolova2004review,nikolova2004fdtd},
or, for S-parameters, from the original fields alone~\cite{nikolova2006td}. Time-domain topology optimization uses the same idea for
antennas and nanostructures~\cite{hassan2014topology,gedeon2025time}, and automatic differentiation (AD) of the whole time-stepping
loop gives the exact derivative of the computed result~\cite{hughes2019forward,kim2026rfx}. In all of these methods, the S-parameters
and their sensitivities are computed from discrete Fourier transforms (DFTs) of responses that stop at a finite simulation time $T$.

In practice, the simulation time is chosen from the response alone. A run continues until the fields have
decayed~\cite{meepsoftware} or the stored energy has fallen by a set amount~\cite{openems}, and the sensitivities are then checked
against finite differences at the same simulation time~\cite{hassan2014topology}. Adjoint formulations integrate over a fixed total
simulation time~\cite{nikolova2004fdtd,zhang2014adjoint}, and the accuracy of time-domain sensitivities computed by the
transmission-line matrix method has been studied against mesh refinement at a fixed simulation time~\cite{nikolova2006td}. In adjoint
design, resonant devices are known to need longer runs for the Fourier-transformed fields to converge~\cite{hammond2022meep}.

The error of an early-stopped response itself is well known. Spectral errors of truncated responses, on filters as elsewhere, have been
studied in microwave time-domain modeling since the early 1990s~\cite{ko1991prony,eswarappa1994prony,kumpel1994si}. The truncated
response has been completed by Prony, system-identification, matrix-pencil, ARMA or dynamic-mode
extrapolation~\cite{ko1991prony,houshmand1993si,cheng2000pencil,shaw2001arma,nayak2024dmd}, or by evaluating the spectrum of the
identified poles directly~\cite{ritter1996pencil}. Resonances and $Q$ have also been estimated from short records by spectrum
estimation, Pad\'e approximation and harmonic inversion~\cite{bi1992fast,dey1998pade,mandelshtam1997harminv}. These methods complete
the response, not its sensitivity.

The convergence of the sensitivity itself has received little attention. Liu and Sarris~\cite{liu2019derivatives} observed that,
on a three-stub microstrip filter, central-difference field derivatives needed 6000 steps where 4000 sufficed for the fields and for
derivatives formed as the product of a separately transformed transfer function and equivalent source.
That a sensitivity converges more slowly is not surprising in itself. When the shift of a pole dominates, the sensitivity of the pole
term is proportional to its frequency derivative, $\partial V/\partial p=-(\partial\omega_r/\partial p)\,\partial V/\partial\omega$,
which is the transform of $t\,v(t)$, and a $t$-weighted transform of a truncated response converges later than the response, as group
delay does. What has not been quantified is how much longer the run must be, and that every exact derivative of the run shares the error. A central difference of two runs is the derivative of the truncated response; in Section~\ref{sec:theory},
we show that this derivative converges later than the response, by a time proportional to the decay time, however the truncated
response is differentiated. A product of separately truncated spectra behaves differently. Each factor is a truncated response, so the
product converges like the S-parameter, but it is not the derivative of the computed S.

Therefore, in this paper, we derive the truncation error of FDTD S-parameter sensitivities, test it, and show what it does to a
design. The contributions are as follows.
\begin{enumerate}
\item A closed-form truncation error of the sensitivity of a damped mode, $(1+T/\tau)e^{-T/\tau}$, and the simulation time it requires
($T\ge 6.6\tau$ for a 1\,\% error), tested in three solvers with no fitted parameter. Because the error belongs to the truncated
computation, AD, the discrete adjoint and finite differences on the same run share it, and their agreement, the usual check of a
sensitivity implementation~\cite{nikolova2004fdtd,hassan2014topology}, cannot reveal it. Earlier sensitivity work fixed
$T$~\cite{nikolova2006td,zhang2014adjoint,hughes2019forward} or observed a slower finite-difference derivative on one
filter~\cite{liu2019derivatives}.
\item The error that common end criteria leave in the sensitivity. These criteria~\cite{meepsoftware,openems}, like the
transit-count rule of~\cite{ritter1996pencil}, watch the response. On a cavity whose slowest mode is weakly coupled, they stop while
$|S_{11}|$ is within 0.5\,dB and the sensitivity is 10--72\,\% in error.
\item Matrix-pencil late-time extrapolation carried inside the differentiable computation, so that the late-time term of the
sensitivity, not only the response~\cite{eswarappa1994prony,kumpel1994si,cheng2000pencil,nayak2024dmd}, comes from the identified
pole. On a cavity with field probes it gives a 1\,\% sensitivity from a run 66 times shorter; on a second cavity, with a longer port
wire, it stays near 2\,\%. A linearized least-squares derivative of
an exponential fit has been taken before on spatial samples, for propagation constants~\cite{swillam2009pencil}; here it acts on the
late time of the S-parameter.
\item The effect on a design, which depends on whether truncation moves the stationary point of the objective. With one pole and a
symmetric objective it does not: an optimizer run at $2\tau$ reaches the resonance of a full run with a gradient up to 71\,\% wrong. With
two coupled poles it does: a loop run for one decay time over-couples by $\Delta k=1.24/Q$, and an equal-ripple objective still shifts
the optimum at $3\tau$. This shift belongs to the truncated objective, and in three solvers it disappears once the S-parameter has
converged. Earlier extrapolation on filters~\cite{eswarappa1994prony,kumpel1994si} corrected the response of a fixed design.
\end{enumerate}
The claims concern the truncation error of the discrete model, so we test them against the same model run to full decay, in three FDTD
solvers (rfx~\cite{kim2026rfx}, Meep~\cite{meep2010}, openEMS~\cite{openems}). The end criteria and the design loops use rfx.

The rest of this paper is organized as follows. Section~\ref{sec:theory} derives the truncation error.
Section~\ref{sec:structures} describes the test structures and solvers, and Section~\ref{sec:estimate} compares the estimate with
simulation. Section~\ref{sec:stop} evaluates common end criteria, and Section~\ref{sec:extrap} presents the late-time extrapolation.
Section~\ref{sec:optimum} asks whether truncation moves the optimum of a design, with one and with two poles.
Section~\ref{sec:discussion} discusses the scope of the results, and Section~\ref{sec:conclusion} concludes the paper.

\begin{figure}[!t]
\centering
\includegraphics[width=\columnwidth]{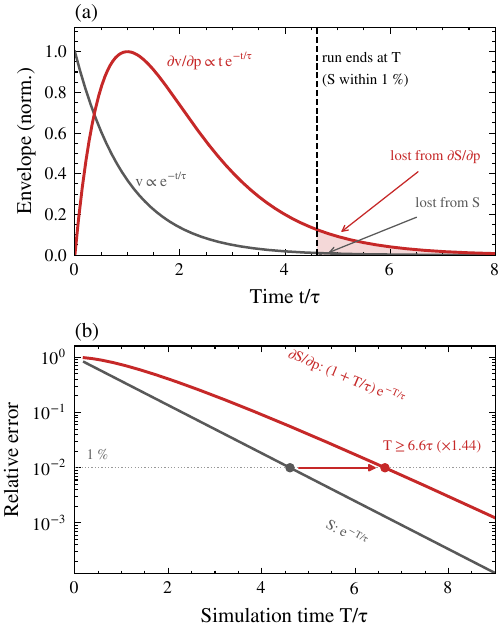}
\caption{Single damped mode with decay time $\tau$ whose pole moves with $p$. (a) Envelopes of the response $v\propto e^{-t/\tau}$ and of its sensitivity
$\partial v/\partial p\propto t\,e^{-t/\tau}$, each normalized to its peak; at resonance $S$ and $\partial S/\partial p$ are their
integrals, and the shaded parts beyond $T$ are what each loses. (b) Relative truncation error of $S$ and $\partial S/\partial p$ at
resonance (lost area over whole area) versus $T/\tau$.}
\label{fig:concept}
\end{figure}

\section{Truncation Error of the Sensitivity}\label{sec:theory}
\subsection{One Damped Mode}
In this section, we derive the truncation error for one damped mode. After the excitation has ended, a port signal is a sum of damped
modes, $v(t)=\sum_k r_k e^{s_k t}$ with $s_k=-\alpha_k+j\omega_k$. The field of mode $k$ decays with the time constant
$\tau_k=1/\alpha_k=2Q_k/\omega_k$, which we use below as the unit of $T$. The sensitivity of one mode to a design parameter $p$ is
\begin{equation}
\frac{\partial v}{\partial p}=\Big(\frac{\partial r}{\partial p}+r\,t\,\frac{\partial s}{\partial p}\Big)e^{st}.
\label{eq:dv}
\end{equation}
When $p$ moves the resonance, the second term grows as $t\,e^{-t/\tau}$. The late-time part of the sensitivity therefore decays more
slowly than the response itself (Fig.~\ref{fig:concept}a). The same $t$-weighting sets the Cram\'er--Rao bound on the frequency of a
tone~\cite{rife1974single} and of a damped sinusoid~\cite{yao1995crb}; here it sets how long the run must be.

In the frequency domain, the DFT of the response from the start of the run, $t=0$, up to $T$ differs from its converged value
$V_\infty=r/(j\omega-s)$ by $V_\infty e^{(s-j\omega)T}$. This is a relative error of $e^{-T/\tau}$ at resonance. Its derivative differs by
\begin{equation}
\frac{\partial (V_\infty-V_T)}{\partial p}
= e^{(s-j\omega)T}\Big[\frac{\partial V_\infty}{\partial p}+V_\infty T\frac{\partial s}{\partial p}\Big].
\label{eq:dV}
\end{equation}
When the shift of the pole dominates, $|(\partial s/\partial p)/(j\omega-s)|\gg|\partial\ln r/\partial p|$, the relative
sensitivity error is
\begin{equation}
\epsilon_g = e^{-T/\tau}\,\big|1+(1+j\Delta\omega\tau)\,T/\tau\big|,\qquad \Delta\omega=\omega-\omega_r .
\label{eq:est}
\end{equation}
At resonance it is $(1+T/\tau)e^{-T/\tau}$, and away from resonance it grows as $|\Delta\omega|T$; $\Delta\omega\tau$ is the detuning in
half-widths (Fig.~\ref{fig:concept}b).

\subsection{The Residue Term and Discrete Time}
The residue term $\partial r/\partial p$, dropped in~\eqref{eq:est}, scales the value at resonance by $|1+\kappa T/\tau|/(1+T/\tau)$, with
$\kappa=1/(1+\alpha(\partial\ln r/\partial p)/(\partial s/\partial p))$. The factor tends to 1 when
$\alpha(\partial\ln r/\partial p)/(\partial s/\partial p)$ is small. High $Q$ favours this but does not guarantee it, and the factor can
lie on either side of 1. Section~\ref{sec:estimate} measures its effect, and Section~\ref{sec:discussion} estimates its size. In discrete time, the factor $(\lambda e^{-j\omega\Delta t})^N$ with
$\lambda=e^{s\Delta t}$ and $N=T/\Delta t$ is exact. Equation~\eqref{eq:est} is its continuous limit, in which $(1-\lambda z)/\Delta t$
also becomes $j\omega-s$.

\subsection{Consequences}
Three consequences follow. First, at the same simulation time, the sensitivity carries $1+T/\tau$ times the relative error of the
S-parameter. For example, a run stopped when the mode has decayed by 40\,dB ($T=4.6\tau$) leaves 1\,\% in the S-parameter of that mode
but 5.6\,\% in its sensitivity.

Second, a sensitivity error of 1\,\% needs
\begin{equation}
T\ \ge\ 6.6\,\tau
\label{eq:rule}
\end{equation}
(6.64 for exactly 1\,\%; we use 6.6), which is 1.44 times the simulation time for a 1\,\% S-parameter. This ratio compares the resonant
part of S. When S also has a nonresonant part, as at a weakly coupled port, S reaches 1\,\% sooner and the gap is wider (Section~\ref{sec:stop}). The
gap is also wider away from resonance: at ten half-widths $\Delta f_{1/2}=\alpha/2\pi$ from resonance, the required simulation time is
1.98 times that of a 1\,\% S-parameter.

Third, AD and finite differences taken on the same run both differentiate the truncated DFT. Their agreement, which is the usual check
of an adjoint implementation~\cite{nikolova2004fdtd,hassan2014topology}, therefore says nothing about $\epsilon_g$.

All three consequences need a parameter that moves the pole (Section~\ref{sec:discussion}). In this paper, $\tau$ is identified on the reference run.
During a design, it would have to come from the run itself, for example from the pole identification of Section~\ref{sec:extrap}.

\subsection{S-Parameters and Notation}
The solver of~\cite{kim2026rfx} forms S from the port voltage and current with a 50\,$\Omega$ reference. On the cavities and the cavity
filters, the excitation pulse ends long before the shortest $T$ used. The incident wave is then settled, and the truncation error of S
is that of the reflected or transmitted wave, to which~\eqref{eq:est} applies. We state errors relative to the whole S entry at the
frequency read. On the notch filter, the pulse lasts $2\tau$ of the nearest pole, so this assumption does not hold at the shortest $T$.

Throughout, $\tau$ is the field (amplitude) decay time $2Q/\omega$ of the feature being read, a half-width is $\Delta f_{1/2}=\alpha/2\pi$
of that feature, and ``a 1\,\% sensitivity'' means a relative error of 1\,\%. In the design loops, $\tau$ is that of the start design
unless stated. $T$ is counted from the start of the run, except on the two-pole filters, where it is counted from the peak of the pulse; on the
cavities the difference is a small fraction of $\tau$. A subscript $T$ ($S_T$, $J_T$, $P_T$) marks a quantity computed from responses
truncated at $T$.

\section{Test Structures and Solvers}\label{sec:structures}
\begin{figure*}[!t]
\centering
\includegraphics[width=\textwidth]{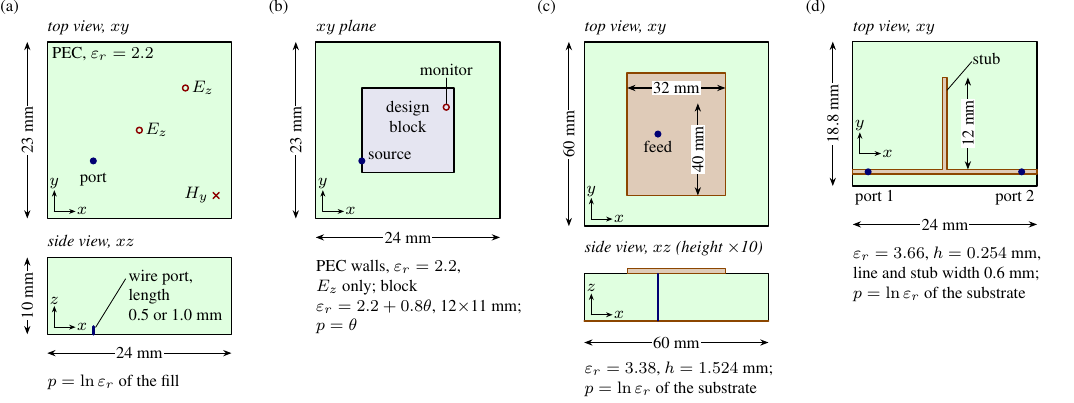}
\caption{Test structures in order of decreasing $Q$ of the feature read, each drawn to its own scale (side view of (c): height $\times10$). Filled dot: port or source; open circle or cross: field probe or monitor (symbols, not to scale). (a) Dielectric-filled PEC cavity read by a wire
port; the three field probes are used only in Section~\ref{sec:extrap}. (b) 2-D cavity of the same footprint in Meep. (c) Patch antenna. (d) Microstrip
notch filter. $p$: design parameter.}
\label{fig:structures}
\end{figure*}
\begin{figure}[!t]
\centering
\makebox[\columnwidth][l]{(a)}\\[-0.2em]
\includegraphics[width=\columnwidth]{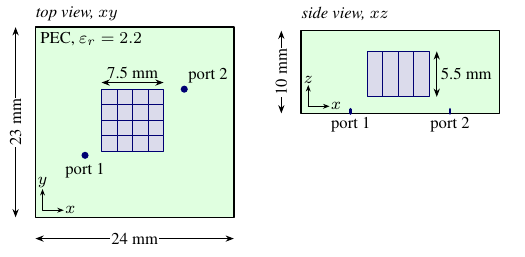}\\[0.4em]
\makebox[\columnwidth][l]{(b)}\\[-0.2em]
\includegraphics[width=0.5\columnwidth]{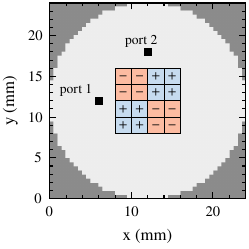}
\caption{Design structures. (a) Single-pole cavity filter: the cavity of Fig.~\ref{fig:structures}(a) with a second wire port and a
dielectric insert of 16 blocks whose permittivities are the design parameters. (b) Circular dual-mode cavity, top view as solved
(dark: PEC), with the sign of $\sigma_b$ on the insert blocks and the two ports.}
\label{fig:design}
\end{figure}

Fig.~\ref{fig:structures} shows the four test structures. We chose them so that the slowest feature ranges from a weakly coupled
high-$Q$ resonance to a transmission zero with no dominant ringing mode.

\emph{Cavity.} A PEC cavity of $24\times23\times10$\,mm filled with $\varepsilon_r=2.2$ is read through a $z$-directed 50\,$\Omega$
wire port at $(6,7.5)$\,mm on the floor. The wire is short, so that dielectric loss, modeled as a conductivity, sets the loaded $Q$ of
TM$_{110}$ at 6.085\,GHz. We use two cavities. With a dielectric $Q$ of 3048 at TM$_{110}$ and a wire 1.0\,mm long, the loaded $Q$ is
$Q_L=2607$ ($\tau=136$\,ns); with 5381 and a wire 0.5\,mm long, it is $Q_L=5025$ (263\,ns). The mesh is uniform with 0.5\,mm cells.
Three field probes, $E_z$ at $(18,17,5)$ and $(12,11.5,5)$\,mm and $H_y$ at $(22,3,8)$\,mm, are used only in Section~\ref{sec:extrap}.

\emph{2-D cavity.} A cavity of the same size in Meep~\cite{meep2010} resonates at 5.735\,GHz, with dielectric $Q_d$ of 300, 1000 and
3000 (mode $Q$ 283, 942 and 2827). Here $p$ is a material weight $\theta$ of a $12\times11$\,mm block at the center, whose permittivity
is $2.2+0.8\theta$.

\emph{Patch antenna.} The patch example of openEMS~\cite{openems} resonates at 2.458\,GHz with $Q\approx 20$ and is fed by a
50\,$\Omega$ wire port.

\emph{Notch filter.} The microstrip notch filter of the openEMS tutorials~\cite{openems}, as carried in the test suite
of~\cite{rfxsoftware}, has its notch at 3.685\,GHz; its nearest pole has $Q\approx 4$.

\emph{Design structures.} Section~\ref{sec:optimum} uses three filters built on such cavities, two of them drawn in Fig.~\ref{fig:design}. The
single-pole cavity filter is the cavity with the 0.5\,mm wire, a second identical wire at $(18,15.5)$\,mm (the point reflection of the
first through the cavity center), the dielectric $Q$ lowered to 2690 at TM$_{110}$, and a dielectric insert of $7.5\times7.5\times5.5$\,mm
divided into 16 blocks whose permittivities are the design parameters. Both wires are weakly coupled (external $Q$ about $7\times10^4$
each), so loss sets $Q_L=2492$ and $S_{21}$ is one loss-dominated pole. The two-pole filters are a square PEC box of $24\times24\times10$\,mm
(dielectric $Q$ about 4160), whose TM$_{120}$ and TM$_{210}$ are degenerate near 9.4\,GHz, and a PEC cylinder of radius 12\,mm and height
8\,mm, whose TM$_{110}$ pair is degenerate near 10.4\,GHz, both filled with $\varepsilon_r=2.2$. As in a dual-mode cavity filter, the two
orthogonal modes of one cavity serve as two coupled resonators~\cite{atia1972narrow,kudsia1992innovations}. A centered insert of
$4\times4$ blocks, $8\times8\times5.5$\,mm in the box and $x,y\in[8,16]$, $z\in[2,6]$\,mm in the cylinder, has
$\varepsilon_r=\bar\varepsilon+u\,\sigma_b$ with $\sigma_b=+1$ on two diagonal quadrants and $-1$ on the other two; $u$ couples the two
modes, and $\bar\varepsilon$ moves both together. Two 50\,$\Omega$ wire ports at $(6,12)$ and $(12,18)$\,mm on the floor, $90^\circ$ apart,
each see one of the modes. The cylinder is solved in all three solvers with the staircase wall of a 0.5\,mm mesh: rfx with wire ports,
openEMS with lumped ports, and Meep with a current source and a port conductance on one edge. Meep uses rfx's edge permittivities cell
for cell; openEMS computes its own by the same averaging rule. The three center frequencies, near 10.399\,GHz, differ by up to
2.5\,MHz, as the dispersion of the Yee scheme at the three time steps predicts, and each solver is read about its own center.

\emph{Solvers and sensitivities.} We simulate the cavity, patch and filter with the AD-enabled FDTD solver of~\cite{kim2026rfx}. The
sensitivity of $S_{11}$ (cavity, patch) or $S_{21}$ (filter) to $p=\ln\varepsilon_r$ of the dielectric, with its conductivity held
fixed, is taken by forward-mode AD on responses truncated at $T$. On the patch, reverse-mode AD, which is the discrete adjoint of the
time stepping~\cite{giles2000adjoint}, gives the same sensitivity to $10^{-5}$ at $T=1$--$4\tau$. In Meep, we differentiate the DFT of a
monitor field by central differences. openEMS~\cite{openems} has neither adjoint nor AD, so we run a 3-D cavity with $Q=1000$ and the
patch tutorial (resonance reproduced at 2.435\,GHz against the documented 2.43\,GHz) at perturbed $\ln\varepsilon_r$, with a step that
moves the resonance by 1--2\,\% of its half-width.

\emph{Converged values.} The converged sensitivity is taken from runs of $16\tau$ or longer; $12\tau$ and $16\tau$ runs agree to
$4\times10^{-4}$. In Meep, the reference runs until the fields have decayed by $10^{-11}$. $\tau$ is that of the feature being read,
identified on the reference run.

\section{The Estimate Against Simulation}\label{sec:estimate}
\begin{figure}[!t]
\centering
\includegraphics[width=\columnwidth]{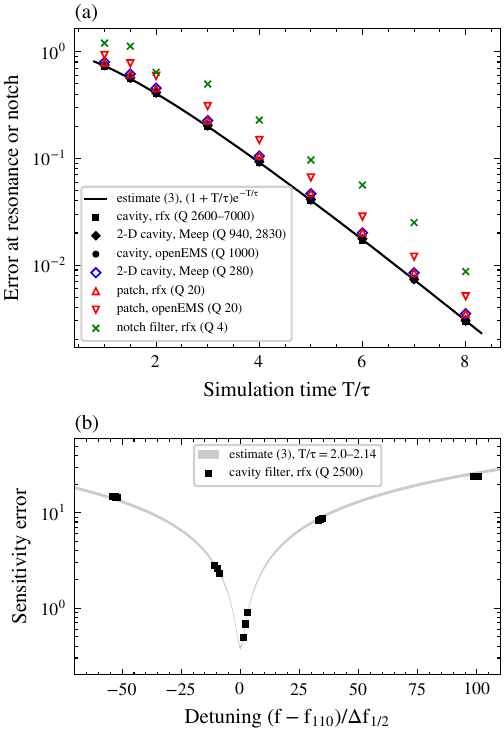}
\caption{Sensitivity error against the estimate~\eqref{eq:est}. (a) At the resonance or notch versus $T/\tau$; filled: $Q$ of about
1000 and above. (b) Single-pole cavity filter at $T=2$--$2.14\tau$ versus detuning from TM$_{110}$.}
\label{fig:law}
\end{figure}

In this section, we compare the estimate~\eqref{eq:est} with the simulated error. The only input of the estimate is the decay rate of
the feature, identified on the reference run. Fig.~\ref{fig:law}(a) shows the simulated error at simulation times of 1 to $8\tau$, at the
frequency point nearest the resonance, where the error is smallest.

Where one mode dominates the response, the simulated error follows the estimate. At cavity resonances with $Q$ from about 1000 to
7000, the ratio of simulated to estimated error stays within 0.98--1.04 in all three solvers. This holds at
TM$_{110}$ on both cavities, at the neighboring TM$_{210}$ and TM$_{120}$ modes, in Meep at $Q_d=1000$ and 3000, and in openEMS at
$Q=1000$ over 1--$5\tau$.

As $Q$ falls, the error keeps the slope of~\eqref{eq:est} but lies above it. It is 7--17\,\% above in Meep at $Q_d=300$ and on the
patch with a wire port, and 26--69\,\% above with the lumped port and excitation of the openEMS tutorial. The residue term dropped
in~\eqref{eq:est} is no longer small there (Section~\ref{sec:discussion}). At the notch, a transmission zero where no single ringing mode dominates and
the pulse lasts $2\tau$ of the nearest pole, the error is 1.6--3.4 times the estimate.

Off resonance, the estimate grows as $|\Delta\omega|T$. Fig.~\ref{fig:law}(b) shows five designs of the single-pole cavity filter,
taken from an optimizer path that moved TM$_{110}$ from 100 half-widths below a target to 55 above, each run fixed at $2\tau$ of the start
design (2.0--$2.14\tau$ of each design), with the sensitivity along the direction that raises $\ln\varepsilon_r$ of all 16 blocks
together, which moves the pole. The error follows~\eqref{eq:est} over the whole range: within 1\,\% of it where TM$_{110}$ dominates
$S_{21}$, and within 11\,\% at 33 to 101 half-widths, where it reaches 24 times the converged sensitivity. A sensitivity read far from the
resonance is therefore wrong by many times its own size, at a simulation time that leaves the on-resonance sensitivity 41\,\% off. In
the derivative of an objective that reads three frequencies, the errors partly cancel near the resonance; far from it, the error is
1.3 to 7 times the derivative itself.

At the resonance or notch, the simulated error is never more than about 2\,\% below the estimate, and off resonance, in these designs,
at most 11\,\% below. The estimate is therefore a prediction at high $Q$ and close to a lower bound elsewhere; because the
residue term can also lower the error, it is not a strict bound.

\section{End Criteria}\label{sec:stop}
\begin{figure*}[!t]
\centering
\includegraphics[width=\textwidth]{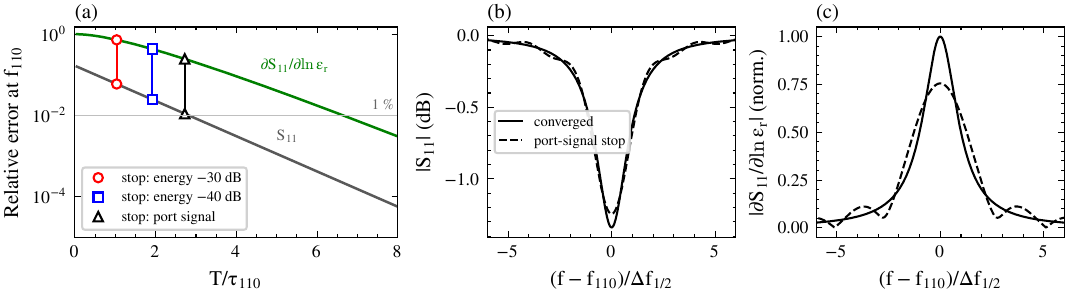}
\caption{Cavity, $Q_L=5025$. (a) Relative error of $S_{11}$ and of $\partial S_{11}/\partial\ln\varepsilon_r$ at TM$_{110}$ versus
$T/\tau_{110}$, with the $-30$ and $-40$\,dB energy and port-signal stops of Table~\ref{tab:stop}. (b) $|S_{11}|$ and (c)
$|\partial S_{11}/\partial\ln\varepsilon_r|$ near TM$_{110}$, converged and at the port-signal stop.}
\label{fig:stop}
\end{figure*}
\begin{table}[!t]
\caption{Error at the Feature When the Run Ends by an End Criterion}
\label{tab:stop}
\centering
\setlength{\tabcolsep}{4pt}
\begin{tabular}{@{}llcccc@{}}
\toprule
structure & criterion & stop & $|S|$ err. & sens. & est. \\
 & & ($T/\tau$) & (dB) & err. & \eqref{eq:est} \\
\midrule
Cavity, TM$_{110}$ & energy $-30$\,dB & 1.04 & 0.50 & 72.3\,\% & 72.3\,\% \\
                   & energy $-40$\,dB & 1.92 & 0.21 & 42.9\,\% & 42.9\,\% \\
                   & energy $-50$\,dB & 2.86 & 0.08 & 22.1\,\% & 22.1\,\% \\
                   & energy $-60$\,dB & 3.85 & 0.03 & 10.3\,\% & 10.3\,\% \\
                   & port signal      & 2.72 & 0.09 & 24.5\,\% & 24.5\,\% \\
                   & DFT, near TM$_{110}$ & 4.40 & 0.02 & 6.6\,\%  & 6.6\,\% \\
                   & DFT, band        & 3.47 & 0.05 & 14.0\,\% & 14.0\,\% \\
Patch              & port signal      & 6.5  & 0.55 & 1.4\,\% & 1.2\,\% \\
Notch filter       & port signal      & 19   & $<10^{-3}$ & $<$0.02\,\% & -- \\
\bottomrule
\end{tabular}
\end{table}

An end criterion decides when the S-parameter has converged. In this section, we apply the three kinds in common use to the
$Q_L=5025$ cavity, whose slowest mode TM$_{110}$ is weakly coupled to the port, and read the error that each leaves in the
sensitivity. One kind watches the field at the ports or monitors, one the energy stored in the whole domain, and one the DFT itself.
Each is applied to stored runs, so that all are read on the same simulation. Fixed times set by the geometry~\cite{ritter1996pencil}
and rules that wait for successive extrapolations to agree~\cite{nayak2024dmd} also watch the response.

\emph{Port-signal criterion.} Meep's field-decay rule~\cite{meepsoftware}, applied to the port voltage, stops when the squared
voltage over ten periods has fallen to $10^{-4}$ of its maximum. On the cavity, the port signal is dominated by faster and more
strongly coupled modes, so the rule stops when TM$_{110}$ has decayed only $2.7\tau$ instead of the $4.6\tau$ that a 40\,dB decay of
that mode needs. On the patch and the notch filter, whose slowest features dominate the port signal, the same rule stops at $6.5\tau$
and $19\tau$, and the sensitivity error is 1.4\,\% and below 0.02\,\% (Table~\ref{tab:stop}).

\emph{Domain-energy criterion.} The second criterion stops when the electromagnetic energy in the whole domain has fallen 30 to
60\,dB below its running maximum; $-50$\,dB is the default of openEMS~\cite{openems}. The energy is the one the Yee
scheme~\cite{yee1966} conserves, $\varepsilon\,\mathbf{E}^n\!\cdot\mathbf{E}^n+\mu\,\mathbf{H}^{n-1/2}\!\cdot\mathbf{H}^{n+1/2}$
summed over cells. Its maximum is set by the excitation, not by the weak mode, and the strongly coupled modes leave first, so the
energy falls faster than TM$_{110}$. The criterion stops after only 1.04 to $3.85\tau$ of TM$_{110}$ (Table~\ref{tab:stop}).

\emph{DFT-convergence criterion.} Meep's decayed-DFT rule~\cite{meepsoftware}, with the DFTs of the port voltage and of $Z_0$ times the
current as monitors, takes the norm of all monitored DFTs about once per period and, after the source has ended, stops when its change
falls to $10^{-11}$ of the largest change seen. The change of the norm is not monotone: single checks fall orders of magnitude below
their neighbors, and the rule stops on the first such check, at 3.47 to $4.40\tau$ depending on the bins monitored
(Table~\ref{tab:stop}), although the ringing of TM$_{110}$ is in those bins.

\emph{What is left.} The energy and port-signal stops leave $|S_{11}|$ at TM$_{110}$ within 0.03--0.5\,dB of its value, where its dip is
1.3\,dB deep, a complex error of 0.4--6\,\%. They leave its sensitivity 10 to 72\,\% in error, 22\,\% at the openEMS default
(Fig.~\ref{fig:stop}, Table~\ref{tab:stop}). The two DFT stops leave 6.6\,\% and 14\,\%. Each sensitivity error equals the estimate at
the stop time within 0.1 percentage point, and the whole truncation curve follows~\eqref{eq:est} (Fig.~\ref{fig:stop}a). The gap
between the errors of $S_{11}$ and of its sensitivity is wider than $1+T/\tau$, because the sensitivity is wholly resonant, while the
resonant part of $S_{11}$ at this weakly coupled port is one sixth of the whole.

In summary, the risk lies at the slowest and most weakly coupled resonance. A criterion that watches the S-parameter or the total
energy stops on the modes that dominate it, not on that resonance. The weak coupling here comes from a short port wire, but the same
holds for any mode weakly coupled to what the criterion watches.

\section{Late-Time Extrapolation in the Differentiable Computation}\label{sec:extrap}
\begin{figure}[!t]
\centering
\includegraphics[width=\columnwidth]{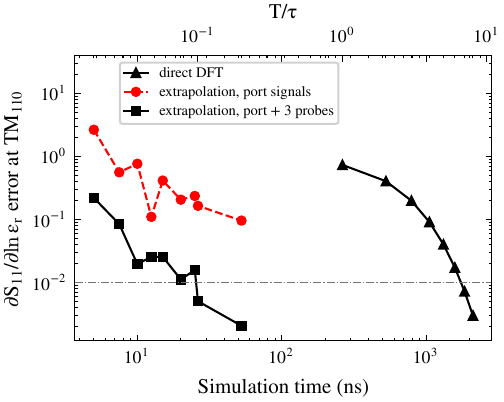}
\caption{Cavity, $Q_L=5025$: relative error of $\partial S_{11}/\partial\ln\varepsilon_r$ at TM$_{110}$ versus simulation time, direct
DFT and late-time extrapolation from the port signals alone or with three field probes (positions in Fig.~\ref{fig:structures}a).}
\label{fig:extrap}
\end{figure}
\begin{table}[!t]
\caption{Simulation Time for a 1\,\% Sensitivity With Late-Time Extrapolation}
\label{tab:cost}
\centering
\begin{tabular}{@{}lccc@{}}
\toprule
case & $\tau$ (ns) & $T$ (ns) & $6.6\tau/T$ \\
\midrule
Cavity, $Q_L=5025$ (rfx) & 263 & 26.3 & 66 \\
Cavity, $Q_L=2607$ (rfx) & 136 & $>$27 & $\le$33 \\
2-D cavity, $Q_d=3000$ (Meep) & 157 & $\le$15.7 & $\ge$66 \\
2-D cavity, $Q_d=1000$ (Meep) & 52.3 & $\le$8.5 & $\ge$40 \\
2-D cavity, $Q_d=300$ (Meep) & 15.7 & $\le$8.5 & $\ge$12 \\
Patch, $Q\approx20$ (rfx) & 2.63 & 5.26 & 3.3 \\
\bottomrule
\end{tabular}
\end{table}

The late-time term $V_\infty T\,\partial s/\partial p$ of~\eqref{eq:dV} belongs to the pole. If the pole and its derivative are
identified from the run, the term can be supplied without simulating the late time. We carry this out inside the differentiable
computation.

The recorded port voltage and current, and optionally field probes, are taken over the window $[T/2,T)$, in which every source waveform
must be below $10^{-6}$ of its peak. Each channel is decimated and scaled by its RMS over the decimated window. The channels are placed
side by side in the Hankel matrices of a matrix pencil~\cite{hua1990pencil} extended to several channels, with pencil parameter one
third of the decimated length. The model order is the number of singular values above $10^{-6}$ of the largest. We discard non-finite
eigenvalues, poles growing by more than $10^{-6}$ per decimated step and poles beyond 0.9 of the decimated Nyquist frequency; nearly coincident poles receive no special treatment. The
residues $c_k$ of each channel are fitted by least squares on the undecimated window, and the response after the last sample $N$ is
added to its DFT as the geometric sum
\begin{equation}
\Delta Y(\omega)=\Delta t\sum_k c_k\,\frac{\lambda_k^{N+1-n_0}z^{N+1}}{1-\lambda_k z},
\label{eq:tail}
\end{equation}
with $\lambda_k=e^{s_k\Delta t}$, $z=e^{-j\omega\Delta t}$ and $n_0$ the first sample of the window. S is then formed from the completed
voltage and current.

For the derivative, we write the poles as $s=s_0+\Delta s$, where $\Delta s$ solves a linearized least-squares system, as
in~\cite{swillam2009pencil}, in the variable-projection form of~\cite{golub1973varpro}, $A\,\Delta s=w\,(Y-\bar Y)$. Here $A$ is the
Jacobian of the RMS-weighted window samples with respect to the poles, projected off the exponential basis (one of the two terms of the
exact variable-projection Jacobian) and frozen at $s_0$, and $\bar Y$ is $Y$ with its derivative removed. In effect,
$\partial s/\partial p=(A^{H}A)^{-1}A^{H}w\,\partial Y/\partial p$: $\Delta s$ is zero in value, and its derivative is the first-order
sensitivity of the least-squares pole fit. The residues are re-fitted at $s$, and AD runs through that fit,~\eqref{eq:tail} and S. The
model order, the discarded poles, the window and the channel scaling are held fixed in the derivative; we call this the frozen
derivative. A check compares the gradient from $[T/2,T)$ with that from $[T/4,T)$ of the same run and accepts it when they differ by
less than $10^{-2}$ of the largest gradient entry. The term $T\,\partial s/\partial p$ of~\eqref{eq:dV} is then supplied by the
identified $\partial s/\partial p$, and the S-parameter and its sensitivity come out of one differentiable computation.

On the $Q_L=5025$ cavity with the three probes, the extrapolation gives the S-parameter within 0.05\,dB and 0.5$^\circ$, and its
sensitivity within 1\,\% of the band's largest sensitivity (0.5\,\% at TM$_{110}$) from 26\,ns, or $0.1\tau$, and within 1.1--2.6\,\%
between 10 and 25\,ns (Fig.~\ref{fig:extrap}). The direct DFT needs 1.6--1.8\,$\mu$s for 1\,\% at TM$_{110}$, so each evaluation is
about 60 to 70 times shorter.

The signals must contain the mode. From the weak port alone, the pole of TM$_{110}$ is found accurately, but the frozen derivative stays
10--42\,\% in error from 12.5 to 53\,ns (Fig.~\ref{fig:extrap}). The port signals still contain the sensitivity, as two other
derivatives show. The first differentiates the pencil itself, with the model order and the selected poles held fixed. The second
identifies the poles afresh on responses perturbed to first order along the recorded derivative, $Y\pm \eta Y'$ with $\eta=10^{-6}$,
and differences the two completed S-parameters; it is defined where both are identified with the same number of poles. Both stay
within 1.6\,\% from 12.5\,ns and agree with each other, so the loss on the port alone lies in the
linearization of the frozen derivative. With the probes, the frozen derivative and the perturbed responses stay within 2.6\,\% from
10\,ns, while the pencil derivative fails at 5 and 20\,ns. We use the frozen derivative, which does not need
the pole count to stay the same.

On the $Q_L=2607$ cavity, whose port wire is 1.0\,mm long, the sensitivity with the probes stays within about 2\,\% from 10\,ns, and
1\,\% is not reached up to 27\,ns. On Meep's monitor field (2-D, one channel), the extrapolated value and finite-difference sensitivity
are within $2\times10^{-5}$ already at the earliest time tested, so Table~\ref{tab:cost} gives an upper bound there. On openEMS port
signals of a dielectric-filled cavity ($Q=1000$), the extrapolation recovers $S_{11}$ to 0.003\,dB at $0.2\tau$. On the patch, the
sensitivity is within 1\,\% at $1\tau$ and $|S_{11}|$ within 0.05\,dB at $2\tau$. The identification uses the voltage and current of
wire ports, so microstrip ports such as those of the notch filter are outside it. The method is part of the open-source
solver~\cite{rfxsoftware}.

Table~\ref{tab:cost} lists, for each case, the time from which the extrapolated sensitivity is within 1\,\% at every simulation time
tested, with the probes on the cavities and with both the value and the sensitivity on target on the patch. Without extrapolation,
every case needs $6.6\tau$ by~\eqref{eq:rule}. Where 1\,\% was met, the extrapolated run took 5 to 26\,ns, set on the patch by the end
of the excitation. The run is about three times shorter on the patch, 66 times shorter on the $Q_L=5025$ cavity, and at least 40 times
shorter on the Meep cavities with $Q_d$ of 1000 and above, where the sensitivity comes from differences of extrapolated runs.

In summary, the late-time term of the sensitivity can be supplied by an identified pole, at a fraction of the run, when the recorded
signals contain the mode.

\section{The Optimum Under Truncation}\label{sec:optimum}
\begin{figure*}[!t]
\centering
\includegraphics[width=\textwidth]{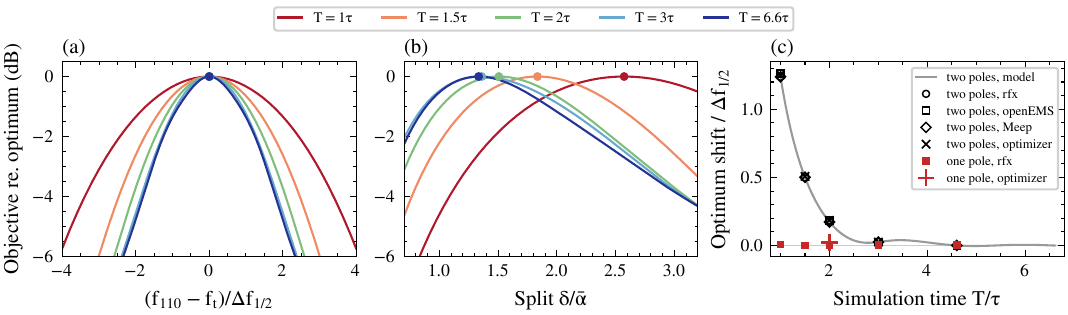}
\caption{Optimum under truncation, from runs of $16\tau$ or longer read at $T$; crosses: optimizer designs from runs fixed at $T$; plus: optimizer run fixed at $2\tau$ against one lengthened by~\eqref{eq:rule}. (a) Single-pole cavity filter: objective versus TM$_{110}$
detuning from the target. (b) Circular dual-mode cavity, rfx: $J_T$ versus the split, as geometric-mean transmission. Each curve is
relative to its own optimum (dot). (c) Shift of the optimum from that at $6.6\tau$.}
\label{fig:pair}
\end{figure*}

A wrong sensitivity does not by itself make a wrong design. A design loop stops where the gradient of the truncated objective
vanishes, so what matters for the design is whether truncation moves that point. We answer this on the two filters of
Fig.~\ref{fig:design}, one with a single pole and one with two coupled poles, in the same way: each design is run for $16\tau$ or longer,
the run is read at several $T$, and the optimum of the truncated objective is located over the design (Fig.~\ref{fig:pair}).

\subsection{One Pole}
The objective of the cavity filter reads $|S_{21}|^2$ at a target frequency $f_t$, with weight 1, and at one half-width on either side,
with weight $1/2$. We shift the permittivity of all 16 blocks together, which moves TM$_{110}$ from four half-widths below the target to
four above in 25 designs. As $T$ shortens, the objective falls and broadens; at $1\tau$ its peak is half that of a long run. Its maximum
stays in place: at every $T$ from 1 to $6.6\tau$ it lies within 0.01 half-width of the target, the resolution of the scan (Fig.~\ref{fig:pair}a,c).

The reason is the symmetry of the truncated spectrum. For one pole, the magnitude of the DFT up to $T$,
$|1-e^{-(\alpha+j\Delta\omega)T}|/|\alpha+j\Delta\omega|$, is even in $\Delta\omega$. An objective that reads the response
symmetrically about the target therefore peaks at the same resonance position whether the run is truncated or not. This needs a
residue nearly constant over the frequencies read, a small direct path between the ports, and a resonance position that moves nearly
independently of the decay rate. All three hold here; with the conductivity held fixed, the decay rate changes by only about $1/Q$ of
the resonance shift.

A gradient loop therefore reaches the right resonance with a wrong gradient. We run the optimizer (Adam~\cite{kingma2015adam},
80 iterations, steps of at most about 0.5 half-width) twice from a start six half-widths below the target, once with the run fixed at
$2\tau$ and once lengthened at every iteration to satisfy~\eqref{eq:rule}. The fixed-run gradient is 3 to 71\,\% in error and above the
on-resonance estimate on most iterations. Evaluated on runs of $16\tau$, the two final designs place TM$_{110}$ within 0.03 half-width of
the target and 0.024 half-width apart (Fig.~\ref{fig:pair}c). The components of the gradient along $Q$ and coupling see a peak that
scales as $(1-e^{-T/\tau})\tau$ instead of $\tau$, and they stay in error at the optimum, with a median error of 63\,\% over the last
iterations.

\subsection{Two Coupled Poles}
A filter is designed by setting couplings, and its passband is shaped by more than one pole. Two poles
$q_\pm=-\bar\alpha+j(\omega_c\pm\delta)$ with residues of opposite sign, a two-pole bandpass response without a zero between the poles,
give the truncated response
\begin{equation}
S_T(\omega)=\sum_\pm r_\pm\,\frac{1-e^{(q_\pm-j\omega)T}}{j\omega-q_\pm}.
\label{eq:twopole}
\end{equation}
With equal decay rates and $|r_+|=|r_-|$, $|S_T|$ stays symmetric about $\omega_c$, so by the argument above the center does not move.
The split does. Truncation broadens each pole by the window, so the passband center is filled less than on a long run, and an
objective that asks for transmission across the passband is satisfied on the truncated response with the poles further apart.

We take the objective $J_T=-\sum_i\ln|S_{21,T}(\omega_i)|^2$ at $\omega_c$ and $\omega_c\pm\bar\alpha$, the geometric mean of the
transmission at the center and one half-width on either side ($\tau=1/\bar\alpha$, $T$ from the peak of the excitation pulse). For two
equal poles, its minimum lies at $\delta^*/\bar\alpha=1.321$ on an untruncated response. Measured from the optimum of a $6.6\tau$ run,
the shortest that gives a 1\,\% sensitivity by~\eqref{eq:rule}, the over-coupling $(\delta^*_T-\delta^*_{6.6\tau})/\bar\alpha$ is 1.238 at
$T=1\tau$, 0.501 at $1.5\tau$, 0.173 at $2\tau$ and 0.026 at $3\tau$. The shift is a property of the truncated objective, not of the
delayed sensitivity: any optimizer of $J_T$ finds it, and a gradient loop finds it because its gradient is the exact derivative of $J_T$,
truncation included.

On the circular cavity, the scan over 18 couplings shows the same move in all three solvers (Fig.~\ref{fig:pair}b,c). The simulated
shifts depart from the prediction re-evaluated from the poles and residues identified on each solver's response by at most
$0.015\bar\alpha$, and from the equal-pole values by at most $0.023\bar\alpha$; the optimum splits agree between solvers within 0.9\,\%.
An optimizer on the square cavity (Adam, 80 iterations, steps of at most about 0.2 half-width in center and split), started at
$\delta=\bar\alpha$ two half-widths below the target center, ends with the center within 0.008 half-width of the target and the shift of its split
within $0.02\bar\alpha$ of the equal-pole shift at each $T$ (Fig.~\ref{fig:pair}c). On a long run, its design from a $1\tau$ run shows a
dip of $-3.44$\,dB at the center of the passband, against $-0.34$\,dB for the design from $6.6\tau$: it is a different filter. In terms of
the coupling coefficient $k=2\delta/\omega_c$ and the loaded $Q=\omega_c/(2\bar\alpha)$ of each pole, the shift at $1\tau$ is
$\Delta k=1.24/Q$.

The over-coupling is not specific to $J_T$. Under the objective used for passband flatness, the largest minimum of $|S_{21,T}|^2$ over
$\omega_c\pm\bar\alpha$, the scanned shifts are 1.32--1.34$\bar\alpha$ at $1\tau$, within $0.035\bar\alpha$ of two equal poles at 1, 1.5 and $2\tau$,
and still $0.18\bar\alpha$ at $3\tau$, where two equal poles give $0.28\bar\alpha$. The equal-pole optimum of this objective passes a kink near $2.7\tau$; read at the
scanned couplings, two equal poles give 0.19.

A run that an end criterion gives is longer than $1\tau$: a domain energy that follows the passband poles falls 30\,dB by about $3.5\tau$.
There, the shift under $J_T$ is below $0.04\bar\alpha$, but under the equal-ripple objective it is still $0.18\bar\alpha$ at $3\tau$ in the scans ($0.28\bar\alpha$ for two equal poles). By
$4.6\tau$, where the resonant part of S is within 1\,\%, the shift is below $0.01\bar\alpha$, the resolution of the scan, in all three solvers
under both objectives, as two equal poles predict (0.001 and 0.002). Runs as short as $1\tau$ arise where the budget per evaluation is
fixed, or where the passband poles hold little of the stored energy, as TM$_{110}$ did in Section~\ref{sec:stop}.

In summary, truncation moves the optimum only where the objective's symmetry does not hold it (Fig.~\ref{fig:pair}c). With one pole, a symmetric objective keeps the optimum in place at
every simulation time, although the gradient is wrong; the error matters where the sensitivity value itself is used. With two poles,
the same symmetry keeps only the center, and the split follows the truncated objective until S has converged.

\section{Discussion}\label{sec:discussion}

\emph{Derivative route.} The delay belongs to the truncated computation, not to the way it is differentiated. Forward- and
reverse-mode AD, the latter the exact discrete adjoint, agree to $10^{-5}$, and central differences on the same run in Meep and openEMS
follow the same estimate (Sections~\ref{sec:structures} and~\ref{sec:estimate}). The term $V_\infty T\,\partial s/\partial p$ in~\eqref{eq:dV} arises because the DFT
stops while the mode it differentiates is still ringing. A frequency-domain adjoint formula behaves differently. It evaluates an
identity derived for converged fields, an integral of forward and adjoint DFT fields over the design region, with both fields truncated
at $T$. Each factor then carries a relative error of order $e^{-T/\tau}$, and no such term appears. Its error is therefore of the value
type, about $2e^{-T/\tau}$ at resonance, one tail per field. It is not the derivative of the computed S, from which it differs by about
$(T/\tau-1)e^{-T/\tau}$ of the sensitivity. A product of separately transformed spectra~\cite{liu2019derivatives} behaves the same way.

Meep's adjoint solver~\cite{hammond2022meep} on the 2-D cavity of Section~\ref{sec:structures}, with a lossless design block (mode $Q$ 3518 and 10554),
confirms this prediction, which was fixed before the runs. At full record, the adjoint agrees with differences within $10^{-4}$. Over
3--$7\tau$, its error follows $2e^{-T/\tau}$ within 4\,\%, and the two derivatives differ as predicted within 3\,\%; at $5\tau$ the adjoint
error is 1.4\,\% against 4.1\,\% for differences ($Q=3518$). The adjoint reaches 1\,\% near $5.3\tau$, where the exact derivative needs
$6.6\tau$. Being the derivative of no computed objective, it can be compared with AD or differences directly only at full
record.\footnote{On a lossy design region, Meep~1.34 returns a field sensitivity divided by $1-j\sigma_D/\omega$: its material gradient
differentiates the inverse permittivity at fixed electric conductivity, while its time stepping holds the loss $-\sigma_DD$ fixed. On a
design block with $\sigma_D/\omega=0.05$, multiplying the adjoint result by $1-j\sigma_D/\omega$ brought it to the finite difference
within $2\times10^{-5}$.}

\emph{Parameter.} The estimate needs a parameter that moves a pole. For one that changes only a residue,~\eqref{eq:dV} reduces to
$e^{(s-j\omega)T}\,\partial V_\infty/\partial p$, the relative error of the S-parameter. All parameters here are permittivities, with the
conductivity held fixed.

\emph{Structure and $Q$.} In units of $\tau$, the delay is the same at every $Q$. In absolute time it grows with $Q$, so it matters for
resonators, filters and narrowband antennas and little for broadband radiators. The estimate is quantitative where one mode dominates.
The residue term scales~\eqref{eq:est} by $|1+\kappa T/\tau|/(1+T/\tau)$, with
$\kappa-1\approx-\alpha(\partial\ln r/\partial p)/(\partial s/\partial p)$. For $p=\ln\varepsilon_r$ of a material holding a fraction $F$
of the electric energy of the mode, $|\partial s/\partial p|\approx\omega_rF/2$, so $|\kappa-1|\approx|\partial\ln r/\partial p|/(QF)$.
At fixed $\partial\ln r/\partial p$ and $F$, the departure therefore grows as $Q$ falls, in the order of Fig.~\ref{fig:law}(a). Restoring
the term gives 0.98 in Meep at $Q_d=300$, with $\partial s/\partial p$ and $\partial\ln r/\partial p$ from central differences of the pole
identification on a later window of the reference runs. At a transmission zero, $S_{21}$ is a near-cancellation of pole terms, each
larger than $S_{21}$ itself. Their truncation errors decay at different rates and need not cancel, so the relative error can exceed the
estimate, up to 3.4 times at the notch.

\emph{Objective.} The error at each frequency holds for any objective built from S there, although errors at several frequencies partly
cancel (Section~\ref{sec:estimate}). What it does to a design depends on the objective. A symmetric objective on one pole keeps the resonance position,
and passband objectives on two poles shift the coupling (Section~\ref{sec:optimum}). On the single-pole filter, an objective read at one
frequency on the skirt, $(P_T(f_e)-t)^2$ with $P_T$ the truncated transmitted power, meets local maxima that the window term
$1-2e^{-T/\tau}\cos(xT/\tau)+e^{-2T/\tau}$ of $P_T$ ($x$: detuning in half-widths) adds to the skirt. On runs of $1\tau$ and $2\tau$, the
optimizer stopped at four such maxima, while a $6.6\tau$ run met the level $t$. In each case the optimizer reaches a stationary point of
the truncated objective, which any optimizer of it can reach. On nearly coincident poles, the late-time sensitivity can decay more
slowly than $t\,e^{-t/\tau}$, and the single-mode estimate does not apply; the two-pole results concern the optimum, not the sensitivity
error.

\emph{Extrapolation in the loop.} Inside the single-pole loop, the two-window check rejected the extrapolated gradient at the first
design, where the gradients from the two windows differed by 1.9\,\% of their peak against the 1\,\% accepted, and the loop ran on the
direct DFT.

\emph{Mesh.} The two-pole shift on the square cavity changes by less than 6\,\% on finer cells; the circular cavity was solved on
0.5\,mm cells in all three solvers. On the single-pole filter, the wire coupling changes with the mesh ($Q_L$ from 2492 to 2660 and
$|S_{21}|$ at the target from $-23$ to $-41$\,dB between cells of 0.5, 0.25 and 0.167\,mm), and the two optimizer designs move together
by about 0.4 half-width. The 0.024 half-width between them compares two runs on one mesh, not an absolute design position.

\section{Conclusion}\label{sec:conclusion}

In this paper, we have studied how the simulation time of an FDTD run limits the accuracy of S-parameter sensitivities. When a
design parameter moves a resonance, the sensitivity converges later than the S-parameter. At resonance, its truncation error is
$(1+T/\tau)e^{-T/\tau}$ for a dominant mode, against $e^{-T/\tau}$ for the resonant part of the S-parameter, so a 1\,\% sensitivity
needs $T\ge6.6\tau$ of the slowest mode it depends on. The error belongs to the truncated computation. AD, the discrete adjoint and
finite differences on the same run share it, so their agreement cannot reveal it. In three solvers, the estimate predicted the
simulated error within 4\,\% at cavity resonances with $Q$ from about 1000 to 7000.

On a cavity whose slowest mode is weakly coupled, common end criteria stopped the run after 1.04 to $4.40\tau$ of that mode. The energy
and port-signal stops left $|S_{11}|$ within 0.5\,dB of a 1.3\,dB dip but its sensitivity 10--72\,\% in error. Either the run must be
long enough, or the late-time term must be supplied in another way; an extrapolation inside the differentiable computation supplied it
from the identified pole, on a run 66 times shorter on one cavity and to about 2\,\% on a second.

The effect on a design depends on the objective. In a single-pole loop with a symmetric objective, truncation left the maximum of the
objective in place, and the resonance was kept despite a wrong gradient. With two coupled poles, minimizing $J_T$ on runs of $1\tau$
over-coupled the resonators by $\Delta k=1.24/Q$, in an rfx optimizer and in coupling scans in three solvers, and an equal-ripple
objective still shifted the optimum at $3\tau$. This shift comes from the truncated objective and disappeared, in all three solvers, once
the S-parameter had converged.

In conclusion, where the sensitivity value itself is used, as in a comparison with a converged or independent reference, a tolerance
or yield estimate, or a step length, the run must be long enough for the sensitivity, $6.6\tau$ of the slowest mode, not only for the
S-parameter. With two coupled poles, the optimum needs a converged S-parameter, $4.6\tau$ for 1\,\%. In future work, we will show the effect on a tolerance
analysis, carry the extrapolation inside the design loop, and extend the analysis to transmission zeros and to filters with more than
two poles.

\section*{Acknowledgment}
The author used Claude (Anthropic) to draft and edit the text of Sections I--IX and to write the figure-generation scripts, and Codex
(OpenAI) to write simulation and post-processing scripts; all content was reviewed, verified against the simulation records, and revised
by the author, who takes full responsibility for it.

\bibliographystyle{IEEEtran}
\bibliography{references}
\end{document}